\documentclass[letterpaper]{article} 
\usepackage{aaai24}  
\usepackage{times}  
\usepackage{helvet}  
\usepackage{courier}  
\usepackage[hyphens]{url}  
\usepackage{graphicx} 
\def\UrlFont{\rm}  
\usepackage{natbib}  
\usepackage{caption} 

\usepackage{comment}
\usepackage{algorithm}
\usepackage{mathtools}
\usepackage{esvect}
\usepackage{comment}
\usepackage{amsmath}
\usepackage{multicol}
\usepackage{tabularx} 
\usepackage{enumerate}
\usepackage{multirow}

\usepackage{booktabs}
\usepackage{algpseudocode}

\usepackage{newfloat}
\usepackage{listings}
\DeclareCaptionStyle{ruled}{labelfont=normalfont,labelsep=colon,strut=off} 
\floatstyle{ruled}
\newfloat{listing}{tb}{lst}{}
\floatname{listing}{Listing}
\title{\title{Estimating Environmental Cost Throughout Model's Adaptive Life Cycle}}

\author {
    Vishwesh Sangarya,\textsuperscript{\rm 1}
    Richard Bradford, \textsuperscript{\rm 2}
    Jung-Eun Kim \textsuperscript{\rm 1}\thanks{Correspondence.}
}
\affiliations {
    \textsuperscript{\rm 1} Computer Science, North Carolina State University\\
    \textsuperscript{\rm 2} Collins Aerospace
}

\begin{document}

\maketitle

\begin{abstract}
With the rapid increase in the research, development, and application of neural networks in the current era, there is a proportional increase in the energy needed to train and use models. Crucially, this is accompanied by the increase in carbon emissions into the environment. A sustainable and socially beneficial approach to reducing the carbon footprint and rising energy demands associated with the modern age of AI/deep learning is the adaptive and continuous reuse of models with regard to changes in the environment of model deployment or variations/changes in the input data. In this paper, we propose \texttt{PreIndex}, a predictive index to estimate the environmental and compute resources associated with model retraining to distributional shifts in data. \texttt{PreIndex} can be used to estimate environmental costs such as carbon emissions and energy usage when retraining from current data distribution to new data distribution. It also correlates with and can be used to estimate other resource indicators associated with deep learning, such as epochs, gradient norm, and magnitude of model parameter change. \texttt{PreIndex} requires only one forward pass of the data, following which it provides a single concise value to estimate resources associated with retraining to the new distribution shifted data. We show that \texttt{PreIndex} can be reliably used across various datasets, model architectures, different types, and intensities of distribution shifts. Thus, \texttt{PreIndex} enables users to make informed decisions for retraining to different distribution shifts and determine the most cost-effective and sustainable option, allowing for the reuse of a model with a much smaller footprint in the environment. The code for this work is available here: \url{https://github.com/JEKimLab/AIES2024PreIndex}

\end{abstract}

\section{Introduction}

Considering the entire development life-cycle of a neural network model, the impact of the training procedure on the environment is substantial, especially with respect to carbon emissions and energy consumption. It would be preferable for a model to be used frequently for a long time ``as is''. However, this is not always feasible. Models must sometimes adapt to a new distribution, environment, or situation - for example, some ground truths might be changed, some data samples might become stale, or some new data samples might need to come into play. A sustainable solution to such situations, ultimately with regard to energy, carbon emissions, and resource consumption, is \emph{reusing} an existing model to adapt to such changes. That is, models can be retrained and adapted to new distributions with minimal retraining instead of training from \emph{scratch}. Retraining can achieve satisfactory accuracy on the new distribution while exhibiting lower computation costs, thus reducing the carbon footprint and energy consumption during the model's development and deployment. The primary objective of this paper is the introduction of a novel metric designed to predict and estimate the environmental resource costs associated with reusing a model for distributional shifts. We provide empirical evidence to show that model retraining significantly lowers compute costs compared to training a new model from scratch. 

As deep learning models become more prevalent in everyday applications, the associated compute demand increases significantly, leading to substantial electricity consumption for model training and inference. This trend has significant social implications, primarily involving the increased release of carbon compounds into the atmosphere. Various works \cite{verdecchia2023systematic, schwartz2019green} have highlighted the importance of considering sustainable and socially conscious practices in AI research and application. \cite{fu2021reconsidering, bannour-etal-2021-evaluating} detail the growing carbon footprint of recent deep learning models for vision and language tasks. Recent research works \cite{10164743, moro2023carburacy} have shifted towards a sustainability-focused approach to model development by being sustainability-oriented instead of performance-oriented. These research works call attention to the ongoing need for deep learning progress to balance energy demands and carbon emissions with societal concerns. Moreover, there is a requirement for reducing carbon emissions, which in turn contributes to lower climate change, thus protecting areas and populations at risk from the impacts of climate change and scarcity of energy.

Hence, we propose a predictive index (\texttt{PreIndex}) to estimate the environmental cost of retraining a model to new changes in the data. \texttt{PreIndex} can be used to estimate the resources that would be expended if a model is retrained to a new distribution. \texttt{PreIndex} quantifies the change and collapse of the class decision boundaries due to the shift in distribution, and also quantifies the shift in the representation as a result of the changing distribution. 
\texttt{PreIndex} requires only a single forward pass of the model, following which it provides a single concise quantitative estimate to compare and predict a model’s retraining cost. A  lower value of \texttt{PreIndex} indicates that fewer resources would be expended when retrained, and vice versa. 

We conduct experiments to validate that \texttt{PreIndex} is an effective estimator of environmental costs such as carbon emissions and energy usage, as well as several other retraining indicators such as epochs, gradient norm, and change in parameter magnitudes. Through extensive experiments over convolutional architectures and also Vision Transformers (ViT), we show that \texttt{PreIndex} is model agnostic and can be used with different architectures without requiring any modifications to the model structure. By leveraging \texttt{PreIndex}, deep learning practitioners and organizations can make informed decisions on deploying models that meet sustainability and resource usage goals.

\section{Related work}
Distribution shifts can occur due to several factors and can be of different types and intensities as seen in \cite{hendrycks2019benchmarking, arjovsky2020invariant, Hendrycks_2021_ICCV}. Augmentation techniques \cite{hendrycks2020augmix, liu2022randommix, zhang2018mixup, kimICML20, lee2020smoothmix} have shown to provide robustness to certain types of distribution shifts, but are computationally heavy and require training a model from scratch. These methods produce robust models for certain distribution shifts, but have marginal improvements on other distribution shifts. Several studies \cite{geirhos2020generalisation, yin2020fourier, ford2019adversarial} have shown that there is a non-uniform improvement in robustness to the different distribution shifts; in some cases, improvement on one type of noise or corruption results in decreased performance on other distributional shifts. Methods using test time adaptation \cite{lim2023ttn, niu2022efficient, goyal2022testtime, wang2022continual} exhibit only minimal improvements in model robustness, rely on batch data, and fail to provide substantial benefits in scenarios with elevated noise levels. If the test time information is insufficient for adapting the model's prediction, these methods fail to provide accurate and confident outputs during inference.

\cite{schwartz2019green} originally coined the concept of Green AI and Red AI, emphasizing how the substantial growth in computational complexity and resource usage of models led to only marginal enhancements in accuracy. They highlight the need for sustainable practices to go hand in hand with performance improvements when developing models. \cite{verdecchia2023systematic} conducted a systematic survey of recent works in sustainable deep learning and showed that energy consumption and carbon footprint are the two most predominant measures to quantify sustainability. Related works \cite{xu2023energy, xu2021survey, GARCIAMARTIN201975, mcdonald-etal-2022-great} tackle the issue of energy consumption of neural networks. \cite{bannour-etal-2021-evaluating, fu2021reconsidering} demonstrate the significant environmental impact of modern deep learning methods due to the carbon footprint associated with training vision and language models. \cite{10.1145/3531146.3533234} shows that training ViTs (transformer-based architectures)\cite{dosovitskiy2021an} emits a considerably greater amount of carbon compared to convolutional networks. Similarly, \cite{strubell2019energy} evaluates the high energy consumption of transformer-based models, thereby leading to higher carbon emissions. \cite{JMLR:v21:20-312, patterson2021carbon} draw attention to the lack of carbon emission reporting in deep learning research. They provide frameworks for carbon emission measurement and documentation with an emphasis on quantifying the sustainability aspects of training models. \cite{schmidt2021codecarbon, anthony2020carbontracker, budennyy2023eco2ai, lacoste2019quantifying} provide tools and frameworks to estimate carbon emissions based on energy usage while training models and emissions from energy generation. \cite{sangarya2023aggregate} explore the impact of individual noise types on model adaptation by using original-noise image pairs.

\cite{agarwal2022estimating, lee2020gradients, huang2021on} show how gradients are an important tool in measuring the difficulty of samples and identifying samples that belong to new distributions. They show how the gradients are steeper and have larger values for difficult data and for out-of-distribution data. Studies such as \cite{stacke2020measuring} use the change in layer representation to study pathology data and focus their work to individual layers of a model to show it correlates to accuracy loss on domain shifts.

Various research studies \cite{10.1145/3121050.3121062, pmlr-v25-aboumoustafa12} demonstrate that commonly used distance measures are not sufficient to be an effective distance metric. \cite{tolstikhin2018wasserstein, pmlr-v70-arjovsky17a, Faber_2021} study the drawbacks of prevalent divergence metrics for specific use cases. \cite{hubert1985comparing} introduced the Adjusted Rand Index for comparing clustering labels, and \cite{santos2009use} illustrate the use of Adjusted Rand Index to evaluate supervised classification and feature selection. \cite{deng2023towards} compares the performance benefits of retraining the entire pre-trained model versus retraining only the feature extractor for classification tasks.
\section{PreIndex}
In this section, we provide a detailed overview regarding \texttt{PreIndex}, which consists of three components -- inverse adjusted rand index, averaged sample representation distance, and noise variance scaling. We introduce each component individually and provide the final formulation of \texttt{PreIndex} at the end of the section.

Table~\ref{tab:notation} provides all notations used in the following subsections.

\begin{table*}
\centering
\vspace{10pt}
\begin{tabular}{llll}
\toprule
\textbf{Symbol} & \textbf{Description} & \textbf{Symbol} & \textbf{Description} \\ 
\midrule
$p$ & Average sample distance & $X_{clean}$ & Clean Image \\ 
$n_s$ & Number of samples & $X_{noisy}$ & Noisy Image \\ 
$n_l$ & Number of layers & $CentM$ & Centroids matrix \\ 

$n_c$ & Number of clusters & $P$ & PDF function \\ 
$WD$ & Wasserstein distance & $Var$ & Variance function \\ 
$h_{rep}$ & Height of filter representation & $h_{img}$ & Height of image \\ 
$w_{rep}$ & Width of filter representation & $w_{img}$ & Width of image \\ 
$Y$ & Class label of a sampple& $\lambda$ & Constant factor \\ 
$r$ & Cluster labels row sum & $t$ & True labels row sum \\ 
$E$ & Normalized Euclidean distance & $W$ & Flattened Weight vector \\ 
$M$ & Model & $R$ & Representation of a sample\\ 
$c$ & Number of classes in the data & $f$ & Number of filters in a layer\\ 
$ActO_l$ & \multicolumn{3}{l}{Activation output function for layer $l$} \\ 
$s$ & \multicolumn{3}{l}{Standard deviation of pixel differences} \\
$\bar{s}$ & \multicolumn{3}{l}{Average deviation over all intensities of specific noise type} \\
$d$ & \multicolumn{3}{l}{Distance per sample at a given layer}\\
\bottomrule
\end{tabular}
\caption{Notation} 
\label{tab:notation} 
\end{table*}

\subsubsection{Adjusted Rand Index for distribution shift}
\label{PreIndex:ari_for_distribution_shift}
We examine the collapse of decision boundaries between classes' data in the representation space as a result of noise. We quantify this change and shift in the decision boundaries by performing clustering on the data representation of the distribution-shifted data. The shift and collapse of the decision boundaries is quantified by obtaining representation data of the entire distributional shift data, followed by clustering on the representation data to generate cluster labels. The cluster labels and the true labels are evaluated to quantify the change in the decision space. Adjusted Rand Index \cite{hubert1985comparing} is useful to assess a clustering algorithm. Adjusted Rand Index (\texttt{ari}) is defined as,

\begin{equation}
\scriptstyle%
\mathtt{ari} = \frac{\sum_{i,j=0}^{n_c} {n_{i,j}\choose{2}} - \Big[\sum_{i=0}^{n_c} {r_{i}\choose{2}} \sum_{j=0}^{n_c} {t_{j}\choose{2}}\Big]/{n_s\choose{2}}} 
{(1/2) \Big[\sum_{i=0}^{n_c} {r_{i}\choose{2}} + \sum_{j=0}^{n_c} {t_{j}\choose{2}}\Big] -  \Big[\sum_{i=0}^{n_c} {r_{i}\choose{2}} \sum_{j=0}^{n_c} {t_{j}\choose{2}}\Big]/{n_s\choose{2}}}
\label{eq:ari}
\end{equation}

where $t$ represents the total count of true labels for each label in the contingency table of true labels vs. representation labels via clustering. $r$ represents the summed values of representation cluster labels in the contingency table. A contingency table in this scenario is a matrix that summarizes the number of samples belonging to the same cluster or having the same label in both clustering scenarios. Here, by `both clustering scenarios,' we refer to the representation-based clustering labels and the true labels. $n_c$ is the number of cluster labels, which is equal to the number of class labels. $n_{i,j}$ represents the value in each entry of the contingency table, which is common to both cluster labels for a given label $i$ and $j$. $n_s$ is the total number of samples.

\texttt{ari} takes the value $0$ for purely random clustering, and $1$ for identical clustering. For our estimator, it is required to have a low value for decision boundaries which are well separated and high value for boundaries which overlap and result in incorrect representation cluster labels. Hence, for our estimator, \texttt{PreIndex}, we take the complement of \texttt{ari} and define it as

\begin{equation}
\scriptstyle%
    inv\_ari = \frac{(1/2) \Big[\sum_{i=0}^{n_c} {r_{i}\choose{2}} + \sum_{j=0}^{n_c} {t_{j}\choose{2}}\Big] - \sum_{i,j=0}^{n_c} {n_{i,j}\choose{2}}}{(1/2) \Big[\sum_{i=0}^{n_c} {r_{i}\choose{2}} + \sum_{j=0}^{n_c} {t_{j}\choose{2}}\Big] -  \Big[\sum_{i=0}^{n_c} {r_{i}\choose{2}} \sum_{j=0}^{n_c} {t_{j}\choose{2}}\Big]/{n\choose{2}}}
    \label{eq:temp}
\end{equation}

\begin{algorithm}[t]
\caption{Representation data's cluster initialization}
\textbf{Input:} Number of samples $n_s$, where $X_i$ is the $i$th sample\\ \phantom{Input: } Number of labels $c$,\\       \phantom{Input: } Model $M$ with representation output ($R$, $Y$),\\ \phantom{Input: } for each sample where $R$ is the representation data \\ \phantom{Input: } and $Y$ is the label\\
\textbf{Output:} Centroids matrix $CentM$, where $CentM_p$ is \\ \phantom{Output: } the $p$th centroid vector
\begin{algorithmic}[1]
\State $CentM \gets \textit{Empty Vector of size c}$
\For{$i \gets 1$ to $k$}
    \State $cur\_centroid \gets \overrightarrow{0}$
    \State $label\_count \gets {0}$
    \For{$j \gets 1$ to $n_s$}
        \State $(R, Y) \gets M(X_j)$
        \If{$Y == i$}
        \State $cur\_centroid \gets cur\_centroid+R$
        \State $label\_count \gets label\_count + 1$
        \EndIf
    \EndFor
    \State $CentM_i \gets cur\_centroid/label\_count$
\EndFor
\end{algorithmic}
\label{alg:algorithm_centroid}
\end{algorithm}

The data representation is obtained from the final convolution layer for a convolutional network' case while from the final dense layer in the last transformer encoder block for a vision transformer's case. To obtain the representation labels by clustering, we use KMeans with 3 different centroid initialization techniques as follows: 
\begin{enumerate}
  \item[1.] Using the original data to obtain centroids as detailed in Algorithm~\ref{alg:algorithm_centroid},
  \item[2.] Initializing by Kmeans++ \cite{arthur2007k}, and
  \item[3.] Initializing by random cluster assignment, and selecting the cluster with least entropy among 20 random initialization seeds.
\end{enumerate}
The above three initialization schemes result in similar clustering labels. We use Algorithm~\ref{alg:algorithm_centroid} due to its computational efficiency as it does not requiring random re-initialization or iterative assignment of centroids, unlike methods such as repeated random cluster initialization and KMeans++. 
Algorithm~\ref{alg:algorithm_centroid} has quadratic runtime with respect to dataset size and number of class labels. However, it is computed only once to initialize the cluster centroids. This is an efficient approach compared to KMeans++, which has been shown to have a super-polynomial run-time starting from initialization to converge in the worst case. In Algorithm~\ref{alg:algorithm_centroid}, we begin by creating an empty vector that has a size equal to the number of class labels, as depicted in line 1. $CentM$ is a vector of size $c$ when initialized, but as the classes' centroids are obtained, each entry in $CentM$ is a vector itself. In the end, $CentM$ is a matrix of size $c$ x Size of flattened representation.

\subsubsection{Average sample representation distance}
\label{PreIndex:per_sample_representation_shift}
In this subsection, we introduce the average sample distance between representations obtained from the original and distribution-shifted sample. The average sample distance is calculated per layer of the model for each data point and then aggregated to provide a single concise scalar value. Wasserstein distance is used to find the distance between probability distributions obtained from the representation of the original sample and the distribution-shifted sample. The two data distributions are used to perform a forward pass and obtain the probability distribution from the activations of each layer.

Algorithm~\ref{alg:per_sample_layer_distance} provides the detailed procedure to obtain the layer distance per sample for a given layer $l$ of a model. In Algorithm~\ref{alg:per_sample_layer_distance}, functions $P$ and $WD$ represent the functions to compute probability density and the Wasserstein distance function, respectively. $l\_{clean}$ and $l\_{noisy}$ are the activation output of all filters in layer $l$ for clean and noisy images, respectively. $l_{clean}$ and $l_{noisy}$ are vectors of size $f$ -- number of filters in the layer $l$. The activation output of each filter is averaged as depicted in lines 5--7 in the algorithm, using $h_{rep}$ and $w_{rep}$, which are the height and width of each filter representation output, respectively. $P_{clean}$ and $P_{noisy}$ are the probability distributions for the clean and noisy samples, respectively.

As a reference, Wasserstein distance is preferred over KL-Divergence, Bhattacharya distance, Jensen-Shannon divergence, and Hellinger distance. Wassertein distance, unlike KL-divergence, is a true distance metric that exhibits symmetry \cite{10.1145/3121050.3121062}, and in contrast to Bhattacharya distance, Wasserstein distance satisfies the triangle inequality \cite{pmlr-v25-aboumoustafa12}. Studies such as \cite{Faber_2021, tolstikhin2018wasserstein, pmlr-v25-aboumoustafa12} highlight why Wasserstein distance is preferred over KL-divergence and its variants, Jensen-Shannon distance, for scenarios where quantifying the exact difference between the distributions has greater importance than measuring the likelihood between distributions. Additionally, \cite{Ocal_2020} demonstrates that Wasserstein distance is effective in capturing the horizontal distances between distributions within the metric space, unlike Hellinger distance.

The process of obtaining the representation distance per sample and per layer is repeated for all samples, and for all layers of the model. The distances for all samples and across all layers are then averaged to obtain the final average sample distance $p$ as follows:
\begin{equation}
\small
    p = \frac{1}{n_s} \frac{1}{n_l} \sum_{i=1}^{n_s} \sum_{k=1}^{n_l} (d_k)_i
\label{eq:predindex_part1}
\end{equation}
where $(d_k)_i$ is the distance between distributions at a given layer $k$ for the $i$th sample, $n_s$ is the total number of samples used, and $n_l$ is the number of layers in the given model.

\begin{algorithm}[t]
\caption{Calculate layer distance per sample}
\textbf{Input:} $ActO_l$: Layer $l$'s output function with $f$ filters, \\
\phantom{Input: } $X_{clean}$: clean sample ; $X_{noisy}$: noisy sample, \\
\phantom{Input: } $ActO_{l-1}$: output function for layer 1 through $l-1$ \\
\textbf{Output:} $d_l$: layer distance  
\begin{algorithmic}[1]
    \State $(Rep_{clean})_{l-1} \gets ActO_{l-1}(X_{clean})$
    \State $(Rep_{noisy})_{l-1} \gets ActO_{l-1}(X_{noisy})$
    
    \State $l\_{clean} \gets ActO_l((Rep_{clean})_{l-1})$

    \State $l\_{noisy} \gets ActO_l((Rep_{noisy})_{l-1})$

    \For{$k \gets 1$ to $f$} 
        \State $l\_{clean}_k \gets \frac{1}{h_{rep} \cdot w_{rep}} \sum_{i=1,j=1}^{h_{rep},w_{rep}} (l\_{clean}_k)_{i,j}$

        \State $l\_{noisy}_k \gets \frac{1}{h_{rep} \cdot w_{rep}} \sum_{i=1,j=1}^{h_{rep},w_{rep}} (l\_{noisy}_k)_{i,j}$ 
    \EndFor

    \State $P_{clean}, P_{noisy} \gets P(l\_{clean}), P(l\_{noisy})$

    \State $d_L \gets WD(P_{clean}, P_{noisy})$
\end{algorithmic}
\label{alg:per_sample_layer_distance}
\end{algorithm}

\begin{figure*}[th]
    \centering
    \includegraphics[width=\linewidth]{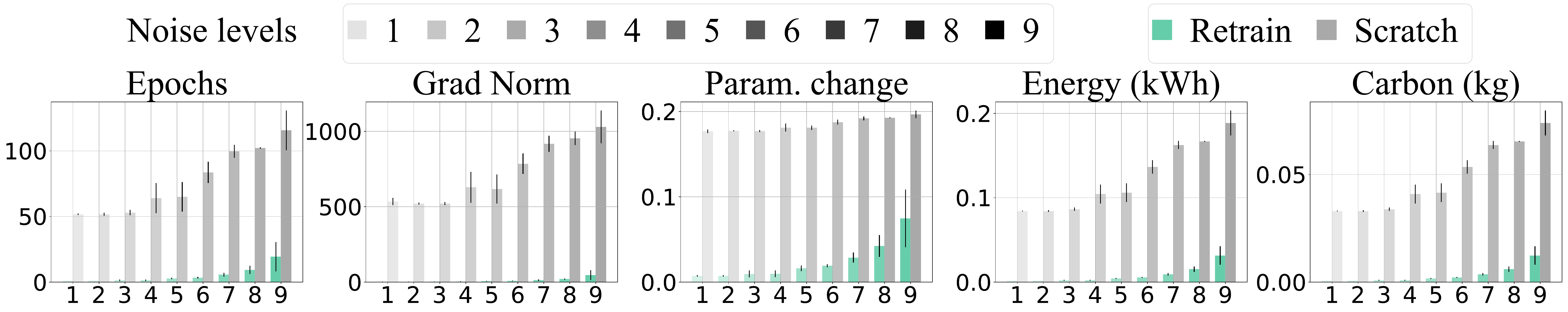}
    \caption{Training from scratch vs. retraining of ResNet18 on CIFAR10 with Poisson noise. Retraining consumes clearly less resources than training from scratch.}
    \label{fig:DistributionShift_ScratchVsRetrain}
    \vspace{-0.1cm}
\end{figure*}

\subsubsection{Noise variance scaling}\label{sec:predindex:noise_variance_scaling}\cite{li2020wavelet, DBLP:conf/ijcai/VargasS20} show that neural networks make incorrect predictions even with small levels of noise in an input image. In particular, \cite{DBLP:conf/ijcai/VargasS20} illustrates the cascading impact of a single pixel change with large magnitude and its effect on neighboring values of the image representation in the deeper layers of a model. 

Different noise types have different traits. In salt-pepper and impulse noises, certain individual pixels (either few or many) are associated with the noise. Hence, the noises affect a specific subset of pixels in an image with a larger magnitude of change per pixel value (that is noised). We refer to this type of noise as pixel-specific noise. Conversely, Gaussian, Blur, Frost, and Poisson noises affect (almost) all pixels in an image with a smaller change per pixel value. We refer to this as global image noise.

Pixel-specific noises, are easier to adapt to since they only affect a subset of pixels as compared to the global image noises. However, the impact of a large magnitude change of a subset of pixels can propagate to the surrounding values in the deeper layer representations. As a result the model may overestimate the raw perturbations caused by pixel-specific noises. To mitigate the overestimation of pixel-specific noises, we introduce an inverse scaling factor. The scaling factor helps reduce the value of \texttt{PreIndex} by utilizing the standard deviation of raw pixel intensities between a clean and a noisy image. We obtain the standard deviation for a specific noise type and intensity as follows:
\begin{equation}
    \small
    s = \sqrt{Var\left([X_{clean(i,j)} - X_{noisy(i,j)}]_{(i,j) \in (h_{img},w_{img})}\right)} / \lambda
        \label{eq:pixel_variance}
\end{equation}
Here, clean image and noisy image are denoted as $X_{clean}$ and $X_{noisy}$, respectively. $h_{img}$ and $w_{img}$ are the height and width of the image, respectively. The resultant standard deviation is then scaled down by a fixed constant factor $\lambda$. $Var$ represents the variance of a given vector.

Finally, \texttt{PreIndex} for quantifying distribution shifts is formulated as,
\begin{equation}
\small
    PreIndex = ( p + inv\_ari )\underbrace{({1}/({1 + (p + (1 - ari)) * s)})}_\text{scaling factor} - \bar{s}
    \label{eq:predindex_stddev_scaled}
\end{equation}

For pixel-specific noises, \texttt{PreIndex} is scaled down using the scaling factor and average deviation, $\bar{s}$. Average deviation $\bar{s}$ is the average of standard deviations across all intensities for the specific noise type. It is used as an offset when scaling down \texttt{PreIndex} for pixel-specific noise. For global image noises, \texttt{PreIndex} is utilized without the scaling factor or average deviation. 
In \texttt{PreIndex}, $p$, $ari$, and $s$ are obtained for each noise type with a specific intensity. Hence, the noise intensity index is omitted for the sake of simplicity.
The values for average sample distance $p$ and $ari$ for each noise type and intensity in Eq.~\eqref{eq:predindex_stddev_scaled} are from Eq.~\eqref{eq:ari} and Eq.~\eqref{eq:predindex_part1}, respectively.
\section{Resource Indicators}
Resource indicators represent the resources/cost that would have been required if a model were trained or retrained to a new target task or to a new distribution. We show that \texttt{PreIndex} has a strong correlation with, and is an effective estimator of, the various indicators listed in this section. Using \texttt{PreIndex} and based on an indicator of interest, a user can gain knowledge regarding the resource expenditure they are likely to expend if they retrain the model to a new task or distribution. We evaluate several indicators - \emph{epochs, gradient norm, change in parameter magnitude, energy, and carbon emissions}. In particular, energy and carbon emissions represent immediate sustainability costs that are likely to embody an ultimate goal to potential users.

\subsection{Epochs}
When a model is retrained and adapted to a new target task or distribution, one might count the training cost/effort by looking at how many epochs are expended. For empirical evidence, we show how consistently \texttt{PreIndex} aligns with the number of epochs, which validates that \texttt{PreIndex} is an effective retraining predictive quantifier. 
The epochs that are reported in this paper are obtained when a model reaches a certain cutoff test accuracy for each dataset. Utilizing additional cutoff conditions, training is terminated after either 25 or 50 epochs if the accuracy gap from the designated cutoff accuracy is within 0.5\% or 1\%, respectively.

\subsection{Gradient Norm}
Gradient norm represents the difficulty of learning a new distribution and provides information regarding the likely steepness to reach convergence. For instance, several studies \cite{agarwal2022estimating, huang2021on, lee2020gradients} use gradient norm as a proxy for sample difficulty. They compute gradients using a uniform distribution and do not make use of ground truth label information. For our objective, we report the gradient norm by utilizing the ground truth label. During retraining, we obtain the overall gradient norm for each instance by aggregating the gradient norm at each time step. The final gradient norm value represents the total magnitude of gradients the model encountered throughout the retraining process.

\begin{table*}[t]
    \footnotesize
    \centering
    \caption{Correlation coefficients (and associated p-value) for multiple models retrained to CIFAR10 against distribution shifts. Results for more datasets and architectures are presented in the Appendix}
    \vspace{-0.3cm}
    \begin{tabularx}{\textwidth}{XXXXX}
    \toprule
    Model & Correlation & Epochs & GradNorm & Param. change \\
    \midrule
    \multirow{2}{\linewidth}{ResNet18} 
    & Pearson & 0.72 (4.2e-10) & 0.66 (4.2e-08) & 0.63 (4.7e-07) \\
    & Spearman & 0.93 (3.1e-25) & 0.93 (1.1e-24) & 0.93 (1.3e-25) \\
    \hline
    \multirow{2}{\linewidth}{GoogleNet} 
    & Pearson & 0.77 (8.0e-12) & 0.73 (1.7e-10) & 0.67 (2.2e-08) \\
    & Spearman & 0.92 (2.8e-23) & 0.91 (5.1e-22) & 0.92 (5.1e-24) \\
    \hline
    \multirow{2}{\linewidth}{VGG16} 
    & Pearson & 0.61 (8.5e-07) & 0.57 (4.9e-06) & 0.55 (1.3e-05) \\
    & Spearman & 0.68 (1.4e-08) & 0.67 (2.4e-08) & 0.67 (2.0e-08) \\
    \hline
    \multirow{2}{\linewidth}{MobileNetv2} 
    & Pearson & 0.76 (1.9e-11) & 0.73 (2.5e-10) & 0.72 (7.4e-10) \\
    & Spearman & 0.81 (4.4e-14) & 0.81 (4.9e-14) & 0.83 (7.7e-15) \\
    \hline
    \multirow{2}{\linewidth}{ViT} 
    & Pearson & 0.75 (5.0e-11) & 0.74 (1.5e-10) & 0.74 (1.4e-10) \\
    & Spearman & 0.81 (5.9e-14) & 0.82 (1.3e-14) & 0.81 (4.3e-14) \\
    \bottomrule
    \end{tabularx}
    \label{tab:noise_correlation}
    \vspace{-0.1cm}
\end{table*}

\subsection{Parameter Change}
Change in parameter value for the entire retraining process represents the magnitude of updates a parameter undergoes. That is, during the retraining process, if the model goes through consistently high parameter changes, it indicates that the model requires further updates and to learn the new distribution.

Hence, it is useful to use \texttt{PreIndex} to estimate a model's parameter change when a model is reused for a new distribution. 

We obtain the change in parameter magnitudes similar to what was done in \cite{zhang2022layers}. However, unlike \cite{zhang2022layers} that calculate changes between parameters' current and initial values (before model updates), we aggregate the change in parameter values between two consecutive time steps. For layer $l$'s parameters, at time step $t$, Normalized Euclidean distance $E_{l, t}$ between present parameter values and parameter values in the previous time step is represented as
\begin{equation}
\small
    E_{l, t} = \left \lVert W_{l,t} - W_{l,t-1}  \right \rVert_2 /\sqrt{\left \lVert W_{l,t}  \right \rVert}
\end{equation}

where $t$ $(\geq 1)$ is the current retraining time step and ${t-1}$ is the previous retraining time step. Distance between current parameter values $W_{l,t}$ and previous parameter values $W_{l, t-1}$ is calculated for each layer $l$, which are then summed and averaged by the number of layers. The normalized Euclidean distance of parameter changes is aggregated throughout the entire retraining process to obtain the final cumulative parameter changes the model undergoes.

\subsection{Energy and Carbon emission}
Energy and carbon emissions provide direct real-world sustainability costs associated with training and/or retraining a model. In accordance with \cite{JMLR:v21:20-312}, reporting carbon emissions is an important sustainability factor when developing models. We use CodeCarbon \cite{schmidt2021codecarbon}, a Python package, to track the energy consumption and estimate the carbon emissions of a model during retraining. The library uses geographic location information of the energy generated to calculate the estimated weight of carbon emissions. The package not only tracks energy consumption from the GPU while training models, but also tracks the energy consumption of the CPU and RAM that is expended by neural network training. The carbon emissions are calculated based on the energy generation technique of the region, such as coal, petroleum, solar, wind etc. 

\section{Experiments}

\begin{figure}[th!]
    \centering
{\includegraphics[width=0.99\linewidth]{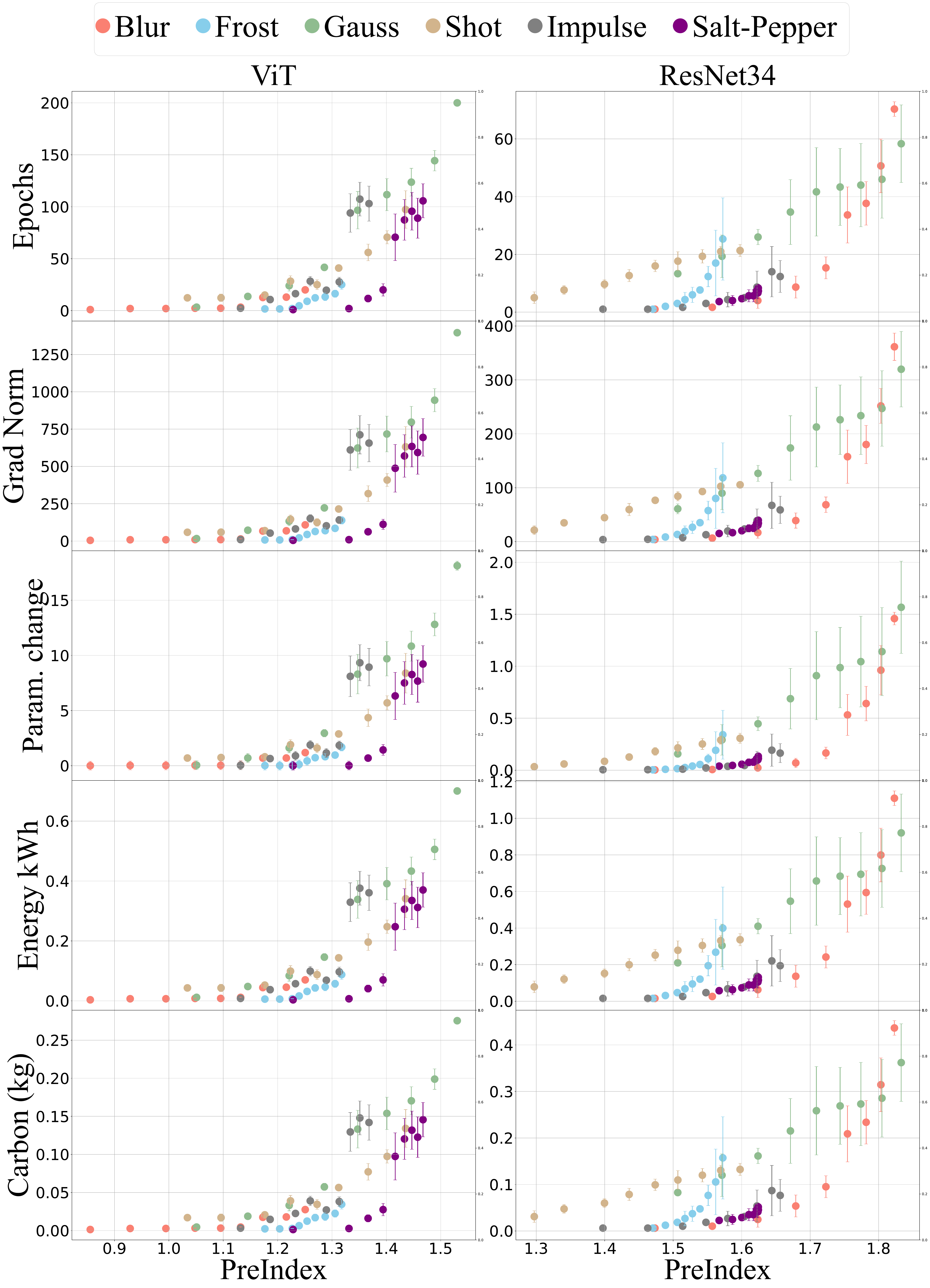}}
    \caption{\texttt{PreIndex} vs. resource indicators on ResNet and ViT. Results of more architectures are presented in the Appendix.}
    \label{fig:noise_resnet34_vit}
    \vspace{2pt}
\end{figure}

In this section, we evaluate \texttt{PreIndex} on three datasets -- CIFAR10 \cite{krizhevsky2009learning}, CIFAR100 \cite{krizhevsky2009learning}, and TinyImageNet \cite{le2015tiny}. We employ 6 noise types - Gaussian, Poisson/Shot, Blur, Frost, Salt-Pepper, and Impulse. Gaussian and Poisson/Shot are statistical noises that may arise due to errors during data capture. Blur and Frost are real-world noises due to environmental factors. Salt-Pepper and Impulse noise might occur due to artifacts or hardware issues while capturing an image. For each noise type, 9 noise intensities are generated, where the lowest level 1 is comparable to the least amount of noise and the highest level 9 is comparable to severity 4 in \cite{hendrycks2019benchmarking}.

\begin{figure*}[t]
    \vspace{5pt}
    \centering
    {\includegraphics[width=\linewidth]{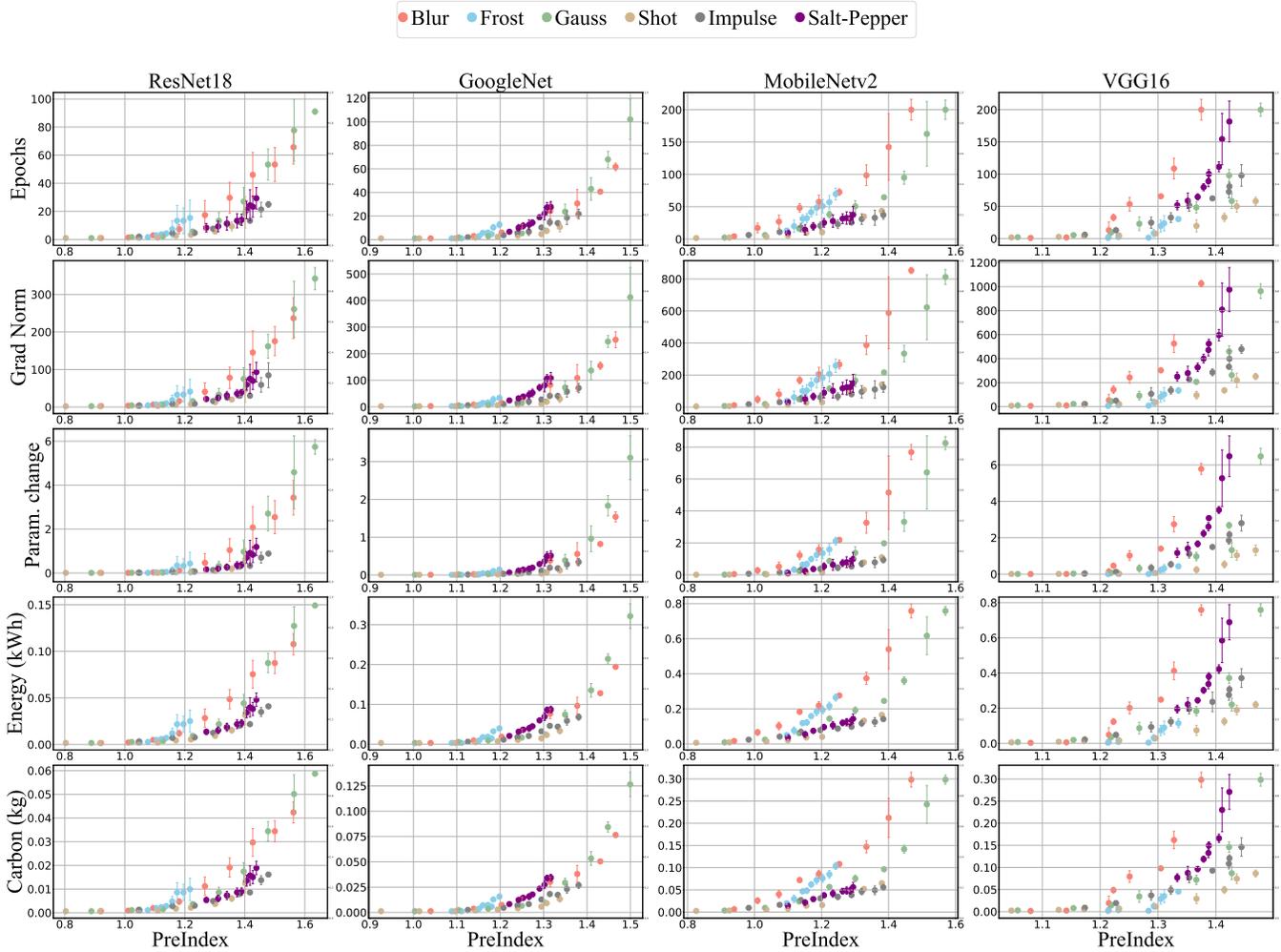}}
    \caption{PreIndex vs resource indicators for various models retrained to CIFAR10 with distribution shift}
    \label{fig:noise_c10}
    \vspace{5pt}
\end{figure*}

To reuse a pre-existing model, we train a randomly initialized model on the original data distribution until it reaches the minimum required accuracy for each dataset. All experiment results are an average of three runs. When training (or retraining) to higher noise levels, specific hyper-parameter tuning for each individual experiment may result in relatively fewer epochs to converge. However, this approach can impede comparison with other noise types and noise levels. For uniformity and fair comparison among all noise types and intensities, identical hyper-parameters are used.

Fig.~\ref{fig:DistributionShift_ScratchVsRetrain} displays the various resource indicators for ResNet18 trained from scratch and retrained on CIFAR10 with different intensities of Poisson noise. It is evident that retraining a model requires considerably fewer resources compared to training a new model from scratch. All the retraining indicator values for retraining are significantly lower than training from scratch.

To evaluate \texttt{PreIndex} for distribution shifts, we use Convolutional neural networks and Vision Transformers. For CNNs, we explore different model architectures - ResNets, VGG, GoogLeNet, and MobileNetv2. For Vision Transformer, we utilize a ViT model with a patch size of 4, 8 transformer blocks, a latent vector size of 512, 8 attention heads, and MLP with a hidden layer size of 1024. 

Fig.~\ref{fig:noise_resnet34_vit} illustrates the correlation between \texttt{PreIndex} and all the resource indicators. In the figure, ViT and ResNet34 are retrained with distribution shifts to CIFAR10 and TinyImageNet, respectively. It is evident that across the heterogeneous architectures, on various datasets, \texttt{PreIndex} has a strong correlation and aligns with all resource indicators for various types of distribution shifts. 

Fig.~\ref{fig:noise_c10} displays the correlation between \texttt{PreIndex} and all resource indicators when retraining convolutional networks to various distribution shifts on CIFAR10 dataset. Table~\ref{tab:noise_correlation} provides the Pearson correlation coefficients and Spearman correlation coefficients, with the associated p-values, between \texttt{PreIndex} and the resource indicators when retraining to CIFAR10 datasets with distribution shift. The figures and the correlation table validate that \texttt{PreIndex} has a strong correlation with the resource indicators, and that it is an effective estimator of retraining resources across various model architectures and types of distribution shifts with multiple intensities.

\begin{figure*}[t]
    \vspace{15pt}
    \centering
    {\includegraphics[width=\linewidth]{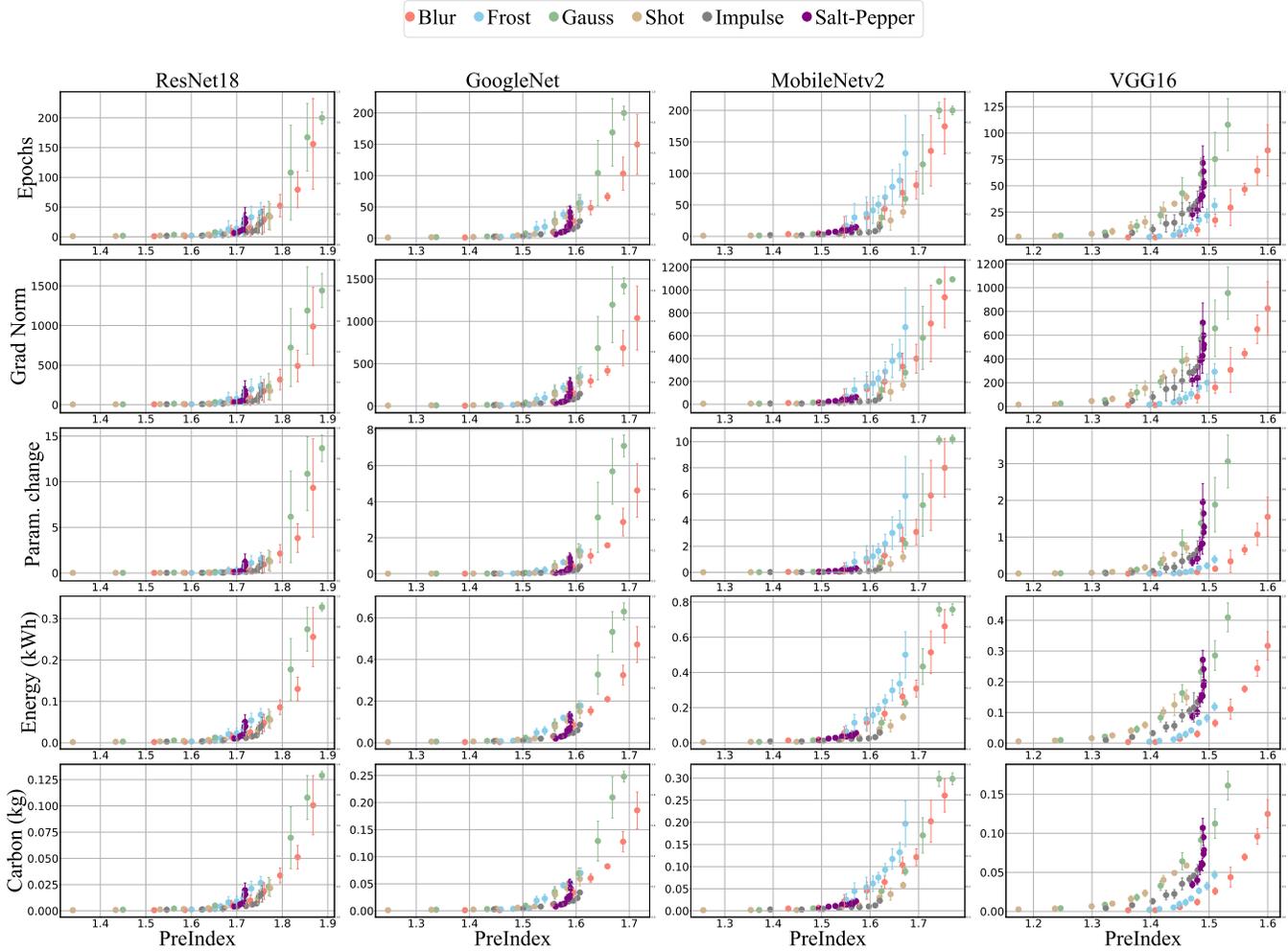}}
    \caption{PreIndex vs resource indicators for various models retrained to CIFAR100 with distribution shift}
    \label{fig:noise_c100}
    \vspace{15pt}
\end{figure*}

\begin{table*}[!h]
    \footnotesize
    \centering
    \caption{Correlation coefficients (and p-values) for multiple models retrained to CIFAR100 with distribution shift}
    \begin{tabularx}{\textwidth}{XXXXX}
    \toprule
    Model & Correlation & Epochs & GradNorm & Param. change \\
    \midrule
    \multirow{2}{\linewidth}{ResNet18} 
    & Pearson & 0.64 (1.5e-07) & 0.62 (6.6e-07) & 0.57 (6.0e-06) \\
    & Spearman & 0.93 (2.4e-24) & 0.93 (7.6e-24) & 0.93 (3.5e-24) \\
    \hline
    \multirow{2}{\linewidth}{GoogleNet} 
    & Pearson & 0.66 (6.8e-08) & 0.63 (3.5e-07) & 0.57 (7.0e-06) \\
    & Spearman & 0.92 (2.1e-23) & 0.92 (3.1e-23) & 0.92 (2.1e-23) \\
    \hline
    \multirow{2}{\linewidth}{MobileNetv2} 
    & Pearson & 0.75 (8.3e-11) & 0.73 (5.1e-10) & 0.70 (5.0e-09) \\
    & Spearman & 0.93 (1.3e-24) & 0.93 (3.0e-24) & 0.93 (5.9e-25) \\
    \hline
    \multirow{2}{\linewidth}{VGG16} 
    & Pearson & 0.67 (3.0e-08) & 0.68 (1.1e-08) & 0.55 (1.7e-05) \\
    & Spearman & 0.81 (1.2e-13) & 0.82 (3.1e-14) & 0.76 (1.6e-11) \\
    \bottomrule
    \end{tabularx}
    \label{tab:correlation_c100}
    \vspace{10pt}
\end{table*}

\begin{figure*}[h]
    \vspace{15pt}
    \centering
    {\includegraphics[width=\linewidth]{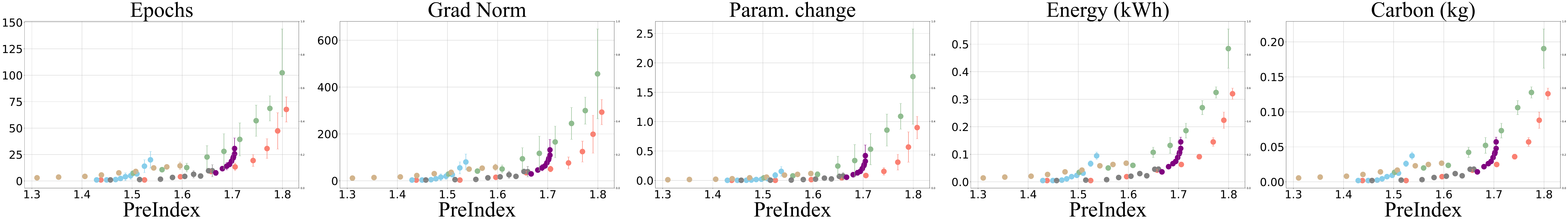}}
    \caption{PreIndex vs resource indicators for GoogleNet retrained to TinyImageNet with distribution shift}
    \label{fig:noise_TinyImageNet}
    \vspace{15pt}
\end{figure*}

\begin{table*}[t]
    \footnotesize
    \centering
    \caption{Correlation coefficients (and p-values) for ResNet34 and GoogleNet retrained to TinyImageNet with distribution shift}
    \begin{tabularx}{\textwidth}{XXXXX}
    \toprule
    Model & Correlation & Epochs & GradNorm & Param. change \\
    \midrule
    \multirow{2}{\linewidth}{ResNet34} 
    & Pearson & 0.73 (1.7e-10) & 0.73 (1.8e-10) & 0.74 (7.7e-11) \\
    & Spearman & 0.62 (4.5e-07) & 0.62 (3.8e-07) & 0.68 (1.2e-08) \\
    \hline
    \multirow{2}{\linewidth}{GoogleNet} 
    & Pearson & 0.70 (2.3e-09) & 0.69 (5.5e-09) & 0.63 (1.9e-07) \\
    & Spearman & 0.83 (3.8e-15) & 0.83 (5.4e-15) & 0.83 (5.1e-15) \\
    \bottomrule
    \end{tabularx}
    \label{tab:correlation_tinyimagenet}
\end{table*}

Fig. ~\ref{fig:noise_c100} and Table~\ref{tab:correlation_c100} provides the trend and correlation metrics between \texttt{PreIndex} and the resource indicators, respectively. With regard to all resource indicators, \texttt{PreIndex} has an increasing monotonic relation noticeable from the graphs and displays a strong positive correlation evident from the correlation coefficients. Additionally, with ResNet18, GoogleNet is evaluated on the TinyImageNet. Fig.~\ref{fig:noise_TinyImageNet} and Table~\ref{tab:correlation_tinyimagenet} illustrate the relation between \texttt{PreIndex} and the resource indicators for GoogleNet retrained on distribution shifted TinyImageNet.

\section{Conclusion}
We introduced a novel metric to estimate the various resources that would be expended when reusing a model by adapting to distributional shifts. We validate the effectiveness of \texttt{PreIndex} on Convolutional and Transformer based networks. We show that reusing a model by retraining requires significantly fewer resources than training a new model. The effectiveness of \texttt{PreIndex} for estimating environmental costs such as energy consumption and carbon emissions, as well as other resource indicators such as epochs, gradient norm, and model parameter change is empirically validated. All the results consistently verify that \texttt{PreIndex} is an effective estimator and has strong correlation metrics with all resource indicators.
\texttt{PreIndex} is shown to be model agnostic, applicable to various datasets and effective for various types and levels of distribution shift, thus enabling sustainable decision making with regard to model reusability.

\bibliography{aaai24}

\end{document}